\documentstyle[emulateapj]{article}

\lefthead{MENOU, HAIMAN \& NARAYANAN} 
\righthead{MERGER HISTORY OF SUPERMASSIVE BLACK HOLES}

\def\gsim{\;\rlap{\lower 2.5pt
 \hbox{$\sim$}}\raise 1.5pt\hbox{$>$}\;}
\def\lsim{\;\rlap{\lower 2.5pt
   \hbox{$\sim$}}\raise 1.5pt\hbox{$<$}\;}

\begin{document}

\title{THE MERGER HISTORY OF SUPERMASSIVE BLACK HOLES IN GALAXIES}

\author{Kristen Menou,\altaffilmark{1} Zolt\'an Haiman,\altaffilmark{2}
and Vijay K. Narayanan}

\affil{Princeton University, Department of Astrophysical Sciences,
Princeton NJ 08544, USA,\\ kristen, zoltan, vijay@astro.princeton.edu}

\altaffiltext{1}{Chandra Fellow}
\altaffiltext{2}{Hubble Fellow}

\begin{abstract}

The ubiquity of supermassive black holes (SMBHs) at the centers of
nearby luminous galaxies can arise from the multiple mergers
experienced by dark matter halos in hierarchical structure formation
models, even if only a small fraction of these galaxies harbor SMBHs
at high redshifts.  We illustrate this possibility using cosmological
Monte Carlo simulations of the merger history of dark matter halos and
their associated SMBHs.  In our most extreme models, in order to
populate nearly every bright galaxy with a SMBH at $z=0$, only a few
percent of the halos with virial temperatures above $10^4$~K are
required to harbor a SMBH at high redshift. This possibility must be
included in studies of the luminosity function and the clustering
properties of quasars.  We predict the number of SMBH merger events
that are detectable by the gravitational wave experiment {\it LISA},
as a function of redshift, out to $z=5$.  Although the event rates can
be significantly reduced in scenarios with rare SMBHs, a minimum of
$\sim 10$ detectable merger events per year is predicted if SMBH
binaries coalesce efficiently. The observed distribution of events
with redshift could yield valuable information on the SMBH formation
process.  If SMBH binaries do not coalesce, we find that at least
several SMBH slingshot ejections probably occurred from $z=5$ to the
present in each galaxy more massive than $\sim 10^{11} M_\odot$ at
$z=0$. Although our results are sensitive to the minimum cooling mass
assumed for the formation of SMBHs, we expect the qualitative
predictions of our models to be robust.

\end{abstract}

\keywords{black hole physics -- cosmology: theory -- quasars: general
-- galaxies: active, nuclei, interactions -- gravitation --
relativity}

\section{Introduction}

There is growing observational evidence for the presence of
supermassive black holes (SMBHs) at the centers of many, if not all,
nearby, weakly-active or inactive galaxies (Kormendy \& Richstone 1995
and references therein; Magorrian et al. 1998; van der Marel 1999).
Early attempts to relate the properties of the nuclear SMBH to those
of the parent galaxy showed the existence of a linear relation between
the SMBH mass and the galaxy luminosity (or that of its bulge
component in the case of disk galaxies; Kormendy \& Richstone 1995;
Magorrian et al. 1998), albeit with a large scatter (of about 0.5 dex
in the ratio $\log[M_{\rm SMBH}/M_{\rm bulge}]$).  Recently, a
substantially tighter correlation between the SMBH mass and the
stellar velocity dispersion of the host galaxy bulge was
reported by Ferrarese \& Merritt (2000) and Gebhardt et al. (2000),
suggesting that the formation and evolution of SMBHs and the bulge of
the parent galaxy may be closely related.

The nuclear SMBHs are most probably ``dead quasars'', i.e., relics of
quasar activity that may have occurred in many galaxies over their
past history (e.g. Lynden-Bell 1967; Soltan 1982; Rees 1990; Richstone
et al. 1998).  At cosmological distances, the presence of SMBHs in
galaxies is inferred from the powerful quasar activity phenomenon,
which is commonly attributed to the radiatively efficient accretion of
gas onto a SMBH of mass $\sim 10^7-10^9~M_{\odot}$ (Salpeter 1964;
Zel'dovich 1964; Lynden-Bell 1969; Rees 1984; Blandford 1992).  The
simplest assumption relating the observations of nuclear black holes
in both nearby and high redshift objects is that a SMBH is present in
every galaxy at any redshift, while only a fraction of these galaxies
are active at any given time.  A number of models based on this
assumption have been constructed over the years to explain the
cosmological evolution of the quasar population with redshift
(e.g. Efstathiou \& Rees 1988; Small \& Blandford 1992; Haehnelt \&
Rees 1993; Haehnelt, Natarajan \& Rees 1998; Haiman \& Loeb 1998;
Haiman \& Menou 2000; Kauffmann \& Haehnelt 2000; Cavaliere \&
Vittorini 2000) and were shown to be consistent with current
observational constraints at both low and high redshift.

In this paper, we show that models in which only a small fraction of
galaxies at high redshift harbor SMBHs are also consistent with the
presence of SMBHs in the bulges of nearly all local, massive
galaxies. This possibility arises because in standard hierarchical
structure formation models, massive halos experience multiple mergers
during their lifetimes.  {This implies that the mass of central SMBHs
at redshift $z=0$ could originate from just a few selected progenitor
halos, rather than having been distributed more uniformally in all the
progenitors}.  Low-frequency gravitational waves, emitted when the
SMBHs at the centers of merged halos coalesce, are detectable to high
redshifts by the planned Laser Interferometer Space Antenna ({\it
LISA}; Folkner 1998). We estimate the event rates detectable by {\it
LISA} in scenarios with rare galactic SMBHs at high redshift and we
suggest that {\it LISA} could prove a useful tool in constraining the
characteristics of the population of SMBHs over a range of redshifts.

The outline of this paper is as follows. In \S2, we briefly describe
the ``merger tree'' Monte--Carlo algorithm used for our calculations,
some of the key tests carried out to check its validity, and the
implementation of two illustrative models in which only a small
fraction of galaxies at high redshift harbor a SMBH. The results for
the evolution of the SMBH occupation fraction and the merger event
rates as a function of redshift are presented in \S3. In \S4, we
discuss some consequences and important limitations of our results.

\section{The Merger Tree Algorithm}

\subsection{Dark Matter Halos}

We follow the evolution with redshift of a Press-Schechter mass distribution
(Press \& Schechter 1974; hereafter PS) of virialized halos at $z=0$, using the
merger tree algorithm described by Sheth \& Lemson (1999).  For the adopted
cosmology (described below), we found that this algorithm was more accurate in
reproducing a PS mass distribution after many successive merger steps toward
high redshift than the algorithm of Somerville \& Kolatt (1999).  In addition,
Sheth \& Lemson (1999) showed that this algorithm accurately reproduces the
higher order moments of the progenitor mass distribution at high redshifts
seen in numerical simulations.  For an assumed PS mass distribution of
virialized objects at redshift $z=0$, the merger tree is constructed by picking
progenitor masses from an appropriate probability distribution at the next
redshift $z+\Delta z$ (following the Extended Press-Schechter formalism,
hereafter EPS; Lacey \& Cole 1993, 1994), where $\Delta z = 0.05$ is the
redshift step of the merger tree. Throughout this work, we adopt a $\Lambda$CDM
cosmology with $\Omega_0=0.3$, $\Omega_b=0.04$, $\Omega_\Lambda=0.7$,
$h_{100}=0.65$, (see, e.g. Bahcall et al. 1999), and a mass power spectrum with
$\sigma_8=0.9$ and $n=1$ (see Eisenstein \& Hu 1999).  We expect our
qualitative conclusions to remain valid in different cosmologies. Note that the
$h_{100}^{-1}$ scaling for masses and the $h_{100}^{-3}$ scaling for volumes
are not shown explicitly below.

We have carried out several tests of the merger tree algorithm,
including comparisons to the PS mass function at various redshifts.
We find that the merger tree algorithm reproduces a PS mass function
to an accuracy better than 0.1 dex till $z=2.5$, 0.2 dex in the range
$2.5 < z < 4$, and 0.3 dex in the range $4 < z < 5.0$.  The
discrepancy between the merger tree predictions and the EPS
predictions for the mass function of halo progenitors grows toward
higher redshifts, and exceeds a factor of $2$ at $z>5$.  Likewise, the
halo merger rates obtained from the tree agree with the EPS rates to
within a factor of two for $0<z<5$.  A convergence test shows that the
redshift step used for the calculations ($\Delta z=0.05$) does not
significantly affect the merger rate predictions: the discrepancy
between trees with $\Delta z=0.01$ and $\Delta z=0.05$ is largest
($\sim 25 \%$) at $z=5$ and reduces to nearly zero at $z=0$.  These
tests guarantee that the mass functions and merger rates produced by
the merger tree are accurate to a factor of two or better in the
redshift range of interest ($z\leq 5$). We chose not to extend our
results beyond $z=5$, where the accuracy of the merger tree degrades
rapidly.  Further details and tests of the implementation of the
merger tree will be given in a separate publication (Haiman, Menou \&
Narayanan 2001, in preparation).

The merger tree we use describes $10^{4}$ halos picked from the PS
mass function at $z=0$, which are subsequently broken up into $\approx
8.8 \times 10^{4}$ halos by $z=5$.  Because of numerical constraints,
the merger tree tracks the merger history of halos over a finite range
in halos mass, which varies with redshift.  The maximum mass at any
redshift corresponds to the most massive halo among the initial
$10^{4}$ halos drawn from a PS distribution at $z=0$ (corresponding to
$M_{\rm max} \approx 8.4 \times 10^{12} {\rm M_\odot}$).  For the minimum
mass, we adopt the redshift-dependent value (e.g. Navarro, Frenk \&
White 1997)
\begin{eqnarray}
\nonumber
M_{\rm min}(z) = &&
9 \times 10^7 {\rm M_\odot} 
\left(\frac{\Omega_0}{0.3}\right)^{-1/2} 
\left(\frac{T_{\rm vir}}{10^4~K}\right)^{3/2} \\
&&\times \left(\frac{1+z}{10}\right)^{-3/2} 
\left(\frac{h_{100}}{0.7}\right)^{-1},
\label{eq:coolmass}
\end{eqnarray}
corresponding to a virial temperature of the halo of $T_{\rm vir}=10^4$~K.
This value represents the minimum halo mass below which, in a metal-free gas,
baryons cannot cool efficiently within the age of the universe at that
redshift, and therefore cannot form a SMBH.  Although this is a reasonable
choice, the precise value of the limiting cooling mass is uncertain (it depends
on the metallicity and ${\rm H_2}$ chemistry; see, e.g. Haiman, Abel \& Rees
2000).

Once a realization of the merger tree has been generated, it can be used
backwards (i.e., from high $z$ to low $z$) to follow the merger history of dark
matter halos which potentially contain black holes.  We estimate the
statistical errors on the various quantities calculated below from the
dispersion in the corresponding quantities among ten independent realizations
of the merger tree with the same parameters.  We find that the statistical
uncertainties are generally small, and are below the expected level of
systematic errors in the merger--tree method.  The merger tree predictions can
be translated into physical units using the equivalent comoving volume that
contains the initial $10^{4}$ halos described by the merger tree.  The PS mass
function predicts $0.579$~halos per Mpc$^3$ at $z=0$, in the appropriate mass
range $M_{\rm min} (z=0)$--$M_{\rm max}$.  Thus, we infer that our merger tree
simulates a fixed comoving volume of $\Delta V=10^4/0.579 =1.73 \times
10^{4}$~Mpc$^3$.

\subsection{Black Hole Seeds And Evolution}

The various models that have been proposed so far for the formation of
SMBHs in individual galactic nuclei (Rees 1984 and references therein;
Quinlan \& Shapiro 1990; Eisenstein \& Loeb 1995; Silk \& Rees 1998;
Selwood \& Moore 1999; Ostriker 2000) generally lack a robust
prediction regarding the presence or absence of a SMBH in a galaxy.
Here, we explore the predictions for two different illustrative
models, in which SMBHs are assumed to be present in only a small
fraction of all the halos that can potentially harbor a SMBH at high
redshift ($z=5$, the initial redshift in the merger tree).  We
demonstrate that: (a) these two widely different models can
nonetheless be consistent with the near ubiquity of central SMBHs in
luminous galaxies at $z=0$, and (b) the detection of low-frequency
gravitational waves expected from the merger of two SMBHs when the
corresponding parent halos merge provides a good tool to constrain the
extent and properties of the population of SMBHs.  Throughout this
paper, we assume that every halo that is modeled by the merger tree
corresponds to a unique galaxy.

Our models are fully specified by the initial ``SMBH occupation
fraction'', $f_{\rm BHi}(z,M_h)$, defined as the fraction of halos of
mass $M_h$ at the initial redshift $z$ (= 5 in our merger tree) that
harbor a SMBH.  In model~I, we assume that SMBHs are randomly seeded
in halos at $z=5$ with an initial SMBH occupation fraction $f_{\rm
BHi}=3 \times 10^{-2}$ (independent of the halo mass above the cooling
threshold):
\begin{equation}
f_{\rm BHi}(z=5,M_h)=
\left\{\matrix{
{3 \times 10^{-2}}
\hfill& M_h\ge M_{\rm min}(z=5)\hfill\cr
0
\hfill& M_h<M_{\rm min}(z=5)\hfill,\cr}\right.
\label{eq:fbhiI}
\end{equation}
Here $M_{\rm min}(z=5)\approx 2\times10^8~{\rm M_\odot}$ is the
cooling mass threshold at $z=5$.

In model~II, the SMBH occupation fraction, averaged over all halos, is $\bar
f_{\rm BHi}= 3 \times 10^{-2}$ as well, but we assume that only the most
massive halos present at $z=5$ are seeded with a SMBH\footnote{We do not
consider a model in which SMBHs are preferentially formed in small mass halos
because their shallow potential wells do not favor such a process (see, e.g.,
Haehnelt et al. 1998; Silk \& Rees 1998).}:
\begin{equation}
f_{\rm BHi}(z=5,M_h)=
\left\{\matrix{
{1}
\hfill& M_h\ge M_{\rm crit}\hfill\cr
0
\hfill& M_h<M_{\rm crit}\hfill,\cr}\right.
\label{eq:fbhiII}
\end{equation}
Here $M_{\rm crit} \approx 5 \times 10^9~{\rm {\rm M_\odot}}$ is the
critical mass such that the overall occupation fraction, averaged over
all halos with masses above $M_{\rm min}(z=5)$, is $3 \times 10^{-2}$.

We follow the evolution with cosmic time of this initial SMBH population, under
the following assumptions: (a) a halo merger remnant contains a SMBH only if at
least one of its progenitors already contained a SMBH prior to the merger, and
(b) SMBHs in a merger remnant coalesce on a timescale that is small
compared to the tree redshift step ($\Delta z=0.05$).  
We discuss the validity of these assumptions in detail in \S3.

In addition, we need to specify the evolution of halos whose masses are close
to the limiting mass $M_{\rm min}$. As mentioned above, we do not track the
merger history of halos whose mass falls below the limit $M_{\rm min}(z)$ given
in equation~(\ref{eq:coolmass}).  As a result, when the merger tree is followed
from high $z$ to low $z$, these halos gradually ``appear'' in the tree. In
model~I, these halos are assigned the same seed fraction $f_{\rm BHi}=3 \times
10^{-2}$, so that they have the same mass-independent probability of containing
a SMBH (irrespective of redshift). On the other hand, model~II assumes that
only the most massive halos at $z=5$ contain SMBHs. Since the newly appearing
halos at $0<z<5$ have small masses by construction in model~II, we assume that
they do not contain a SMBH.

Note that the masses of SMBHs are left unspecified in our models.  This
omission, which greatly simplifies the problem, is motivated by the fact that
{\it LISA} can, in principle, detect the gravitational waves emitted in any
merger involving black holes with individual masses $> 10^{3} M_{\odot}$ up to
very high redshifts (see \S~3.3.1 below). The accretion history of SMBHs, and
therefore the evolution of their masses, will be explored in a future paper
(Haiman, Menou \& Narayanan 2001, in preparation).  In the present work, we
focus only on the presence or absence of a SMBH in any given halo, as a
function of redshift.

\section{Results}

\subsection{Black Hole Occupation Fraction}

Figures 1a and 1b show the evolution with redshift of the black hole
occupation fraction $f_{\rm BH}$ in model~I and model~II,
respectively.  The three panels in each figure show, from top to
bottom, the black hole occupation fraction for halos with masses
$M_{h} > 10^{11} {\rm M_\odot}$, $ M_{h} > 10^{10} {\rm M_\odot}$, and
for all the halos described by the merger tree (corresponding to the
mass range $M_{\rm min}(z) < M_{h} < M_{\rm max}$).  In both models,
the value of $f_{\rm BH}$, averaged over all halos above the limiting
mass, is $\bar f_{\rm BHi} \equiv 3 \times 10^{-2}$ at $z=5$.
However, the difference in the initial conditions in the two models
(Eqns.~[\ref{eq:fbhiI} and \ref{eq:fbhiII}]) strongly influences the
subsequent evolution of $f_{\rm BH}$ via the merging process.
Overall, $f_{\rm BH}$ computed over the entire mass range $M_{\rm
min}(z) < M_{h} < M_{\rm max}$ increases with cosmic time because of
the numerous mergers experienced by the halo population.  However, the
evolution of $f_{\rm BH}$ depends strongly on the halo mass.  For
example, in model~II, only the most massive halos (with $M_{\rm h}
\gsim 5 \times 10^{9} {\rm M_\odot}$) are seeded with a SMBH at $z=5$,
so that $f_{\rm BH}$ for halos with masses $M_h > 10^{10} {\rm
M_\odot}$ actually decreases with cosmic time because many lower mass
halos that lack SMBHs become more massive than $10^{10} {\rm M_\odot}$
by successive mergers.  This decrease in $f_{\rm BH}$ for $M_h >
10^{10} {\rm M_\odot}$ is opposite to the evolution seen in model~I,
where the successive mergers of small halos, some of which contain
SMBHs, increase the value of $f_{\rm BH}$ with cosmic time for halos
with masses $M_h > 10^{10} {\rm M_\odot}$.

Another important effect of the mass-scale dependence of the merging process is
seen in the value of $f_{\rm BH}$ for the most massive halos.  In both model~I
and~II, halos with masses $> 10^{11} {\rm M_\odot}$ (corresponding roughly to
the masses of galaxies brighter than $0.1L^{*}$ at $z=0$) 
have $f_{\rm BH} \gsim 0.8$ at $z=0$.  This
large SMBH occupation fraction at $z=0$ can be understood by observing that
massive halos possess a larger number of early progenitors, and hence the
massive halos at $z=0$ very often had a progenitor seeded with a SMBH (in
model~I).  In addition, massive halos are also more likely than smaller mass
halos to have relatively large mass progenitors, and hence are less likely to
be ``contaminated'' by numerous small mass halos without SMBHs (in model~II).

We also constructed models with values of the parameter $f_{\rm BHi}
=10^{-1}$, $10^{-2}$ and $10^{-3}$ at $z=5$. As might be expected, for
$f_{\rm BHi} =10^{-1}$, the resulting values of $f_{\rm BH}$ at $z=0$
are systematically larger, and nearly unity for the most massive ($M_h
> 10^{11} {\rm M_\odot}$) halos in models~I and~II. For $f_{\rm BHi}
=10^{-3}$, however, the resulting occupation fraction $f_{\rm BH}$ at
$z=0$ for halos of mass $M_h > 10^{12} {\rm M_\odot}$ is only $\approx
40 \%$ and $\approx 60 \%$ in models~I and~II, respectively (and even
less for halos of mass $M_h > 10^{11} {\rm M_\odot}$).  For $f_{\rm
BHi} =10^{-2}$, the resulting occupation fraction $f_{\rm BH}$ at
$z=0$ for halos of mass $M_h > 10^{11} {\rm M_\odot}$ is only $\approx
50 \%$ in both models. Given that the study of Magorrian et al. (1998)
probes several galaxies down to luminosities $\approx 10^{10} {\rm
L_\odot}$ (roughly corresponding to a total mass of $10^{11} {\rm
M_\odot}$ for a mass to light ratio of 10),\footnote{The very few
galaxies with luminosities $\ll 10^{10} {\rm L_\odot}$ in the
Magorrian et al. sample cannot be used as a stastically significant
constraint for our models. They show, however, that SMBHs are present
in at least a small fraction of lower mass galaxies.}  the models
with $f_{\rm BHi} =10^{-2}$ do not satisfy the observational
constraints. Conversely, scenarios with a black hole occupation
fraction $f_{\rm BH} \sim 3 \times 10^{-2}$ or more at $z=5$ are
consistent with the observational evidence for the presence of a SMBH
in nearly all luminous galaxies at $z=0$.

\begin{figure}
\plottwo{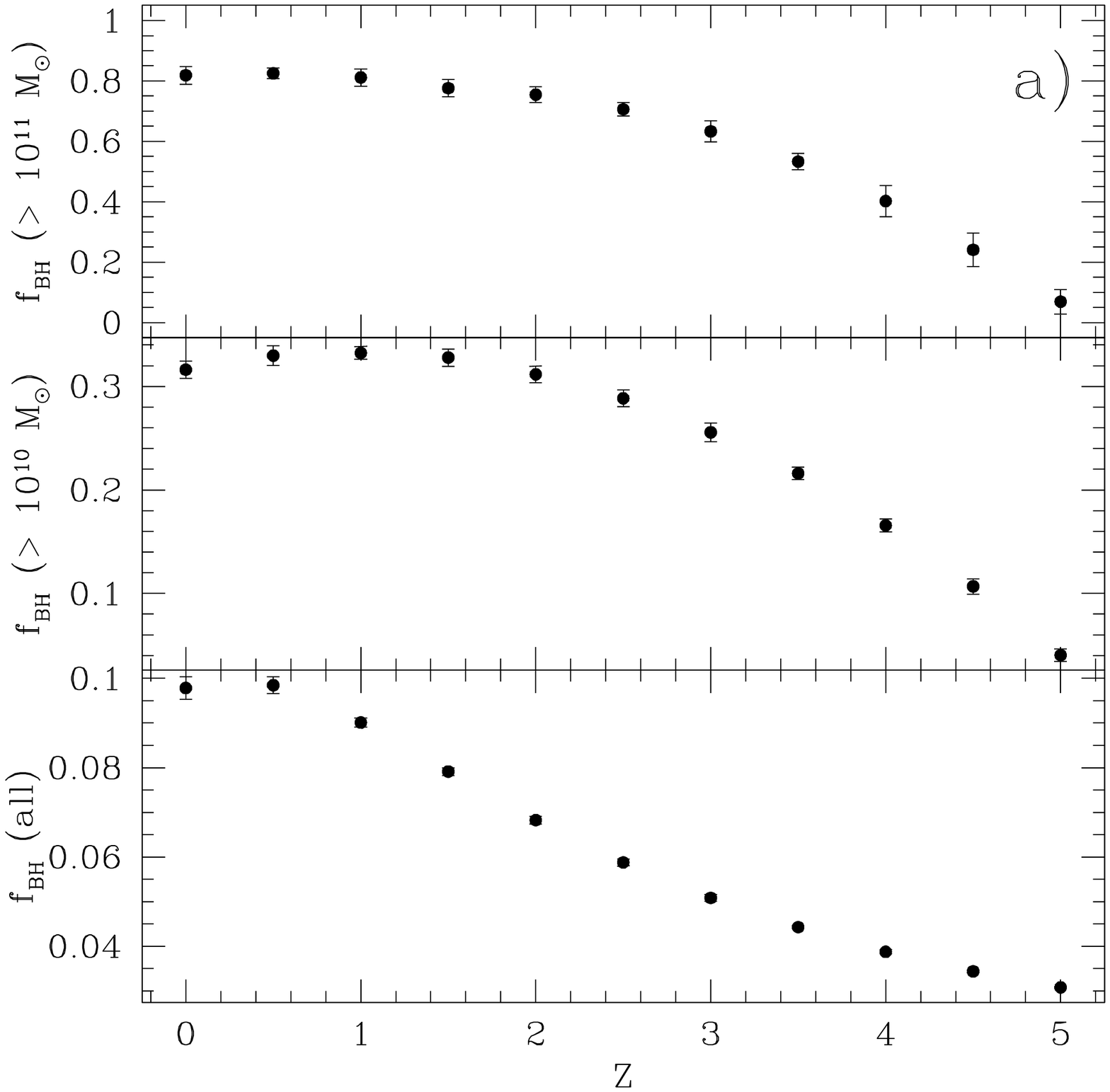}{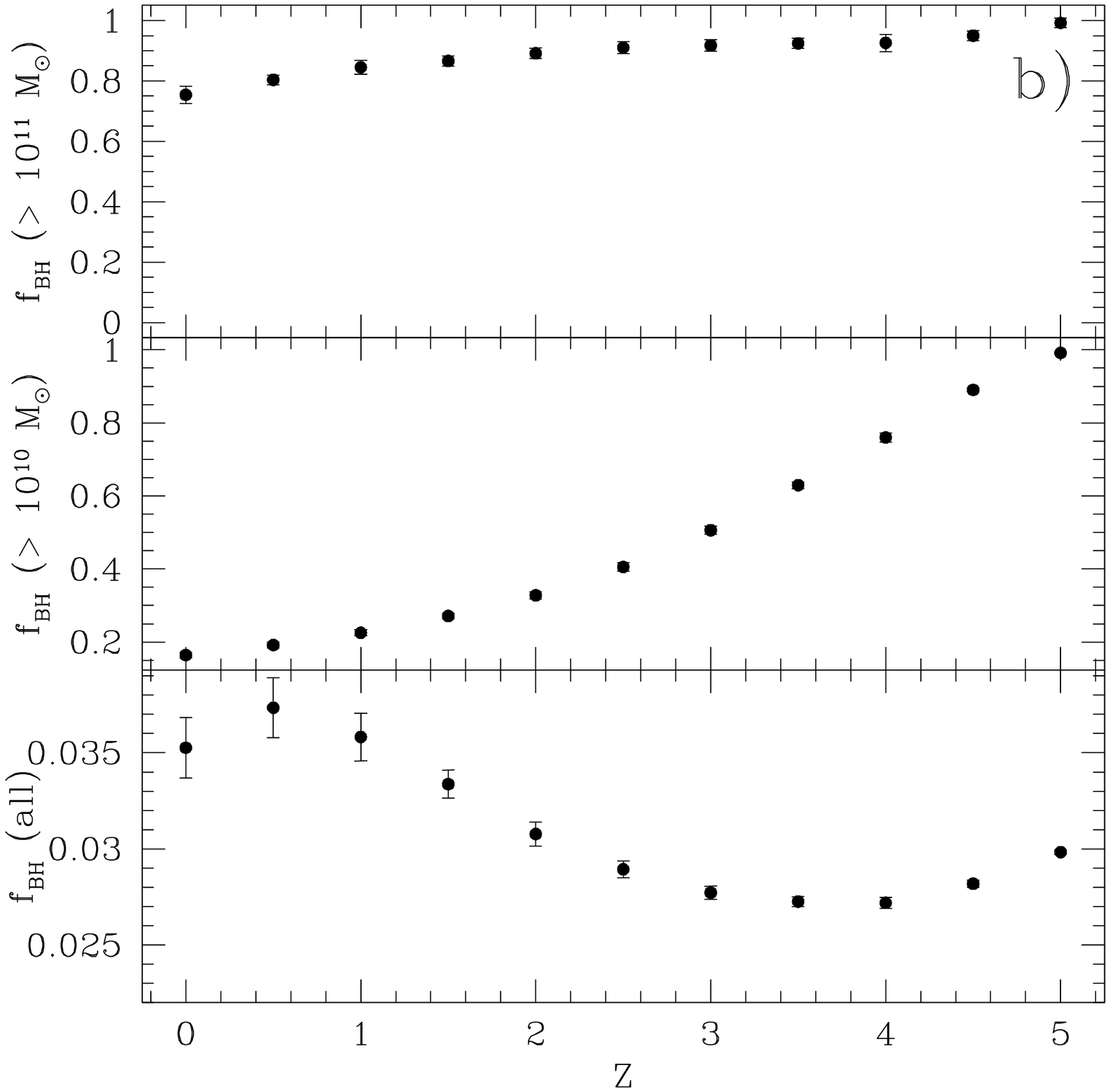}
\caption{The black hole occupation fraction $f_{\rm BH}$ is shown as a
function of redshift in model~I~(a) and model~II~(b). In each case,
the three panels (top to bottom) correspond to halos with mass $>
10^{11} {\rm M_\odot}$, $> 10^{10} M_\odot$ and all the halos (meaning
roughly $\gsim 10^8 {\rm M_\odot}$) described by the merger tree. The large
mass halos ($M>10^{11} {\rm M_\odot}$) nearly all harbor a SMBH at $z=0$
despite the low overall value of $f_{\rm BH}$. Standard deviations are
indicated by error-bars.
\label{fig:one}}
\end{figure}

{One can ask whether the models are consistent with observational
constraints from high redshift quasars (see also \S4 for a discussion
of possible additional constraints on the models). At $z \sim 3$, the
epoch of maximum activity for bright optical quasars,} only a fraction
$\sim 10^{-3}$ of all bright galaxies shows quasar activity
(e.g. Richstone et al. 1998).  This extra constraint is not violated
in the models discussed above.  Similarly, the discovery by Barger et
al. (2001; see also Mushotzky et al. 2000) that a fraction $\sim 10
\%$ of all bulge-dominated optically luminous galaxies at $z < 3$ show
hard X-ray activity does not rule out the models presented.  In fact,
we find a large value of $f_{\rm BH} \gsim 0.7$ for luminous galaxies
at $z \sim 3$ in both classes of models considered (see
Fig.~\ref{fig:one}).  This result follows from our assumption that
SMBHs form mostly (in model~I) or exclusively (in model~II) at high
redshift $z\geq 5$, and that a large fraction $\gsim 80\%$ of halos
with $M_h\geq 10^{11}~{\rm M_\odot}$ at $z=0$ must harbor SMBHs.  In
order to satisfy the constraint at $z=0$, at least $\sim 3\%$ of halos
near the cooling threshold at $z=5$ must have SMBHs; in model~I, the
numerous mergers between halos in the redshift range $3<z<5$ ensures
that a fair fraction of all luminous galaxies must harbor a SMBH by $z
\sim 3$.  This is an important result that shows that a fair fraction
of all luminous galaxies must harbor a SMBH by $z \sim 3$ if SMBHs
preferentially form at redshifts $z>5$.  The rapid decline of quasar
activity beyond redshifts $z \sim 3$ (e.g. Fan et al. 2001) also
allows for scenarios in which SMBHs preferentially form between $3 < z
< 5$. We did not explore this class of models here because the
predictions would crucially depend on the assumed distribution of
formation redshifts for the SMBHs. We note, however, that in a
scenario with an epoch of formation peaking around $z \gsim 3$, we
find that it is possible to have a value of $f_{\rm BH}$ as low as
$\sim 5 \times 10^{-2}$ at $z=3$ and still satisfy the constraints at
$z=0$. This type of scenarios, which would probably lead to reduced
merger event rates between $3 < z < 5$, are better explored with a
simultaneous attempt at reproducing the observed evolution of the
quasar luminosity function (Haiman, Menou \& Narayanan, in
preparation).

The possibility of a relatively rare population of SMBHs at high
redshifts should be taken into account in studies of the evolution of
the quasar luminosity function. In terms of a fit to a luminosity
function, reducing the fraction of galaxies that host SMBHs and go
through a quasar phase can be compensated by either decreasing the
mass ratio between the SMBH and its host halo by a similar factor,
and/or assuming a longer activity phase (i.e. a larger duty cycle).
Studies of the clustering properties of quasars will also have to
account for this possibility: the clustering would be reduced in
scenarios with $f_{\rm BH} \ll 1$ relative to scenarios with $f_{\rm
BH} \sim 1$ at high redshift (Martini \& Weinberg 2000; Haiman \& Hui
2000). The problem could also be complicated by the fact that in
scenarios with rare SMBHs at high redshift, the clustering study will
have to take into account not only the small value of $f_{\rm BH}$ but
its mass--dependence as well.  These issues will be discussed further
in a future paper (Haiman, Menou \& Narayanan 2001, in preparation).

The general trend shown by the models at $z=0$ (due to the mass-scale
dependence of the merging process) corresponds to larger values of
$f_{\rm BH}$ for larger halo masses. The transition from halos mostly
populated with SMBHs to halos without SMBHs occurs over a finite mass
range, which is characteristic of the extent of the population of
SMBHs at high redshift. For example, in both model~I and~II, {we
find that this transition, defined by $f_{\rm BH}=0.5$ (at $z=0$),
occurs at the halo mass} $M_h \approx 3 \times
10^{10} M_\odot$ (see also Fig.~\ref{fig:one}), which is approximately
three times less than the masses extensively probed by the study of
Magorrian et al. (1998).  This property suggests that one way
of constraining the properties of this population is to search for
SMBHs in nearby dwarf galaxies with masses smaller than have been
studied so far (Kormendy \& Richstone 1995; Magorrian et al. 1998).

\subsection{Rates of Black Hole Mergers}

In this section, we show that the detection of gravitational waves emitted when
two SMBHs merge is an efficient technique for determining the extent of
the population of SMBHs over a wide range of masses and redshifts.  It is
possible to calculate the cosmic rate of halo mergers containing SMBHs from the
merger tree. If the corresponding SMBHs coalesce (see \S3.3.2), these
merger events
are potentially detectable by a low-frequency gravitational wave experiment
such as {\it LISA}.  The total number of events $\Delta N$ occurring in the
tree during one redshift step ($\Delta z=0.05$) corresponds to $\Delta N/\Delta
z/\Delta V$ events per Mpc$^3$ per unit redshift (used here as a time
coordinate). Given the adopted cosmology, this event rate can be translated
into the number of detectable merger events per unit time per unit redshift
(used here as a space coordinate; see e.g., Kolb \& Turner 1990; Haehnelt 1994
for similar derivations):
\begin{equation}
\frac{d N}{dz dt}=\frac{\Delta N}{\Delta z~\Delta V} {4 \pi c (1+z)^2
d_A^2(z)},
\end{equation}
where $d_A(z)$ is the angular diameter distance and $c$ is the speed
of light.

\begin{figure}
\plottwo{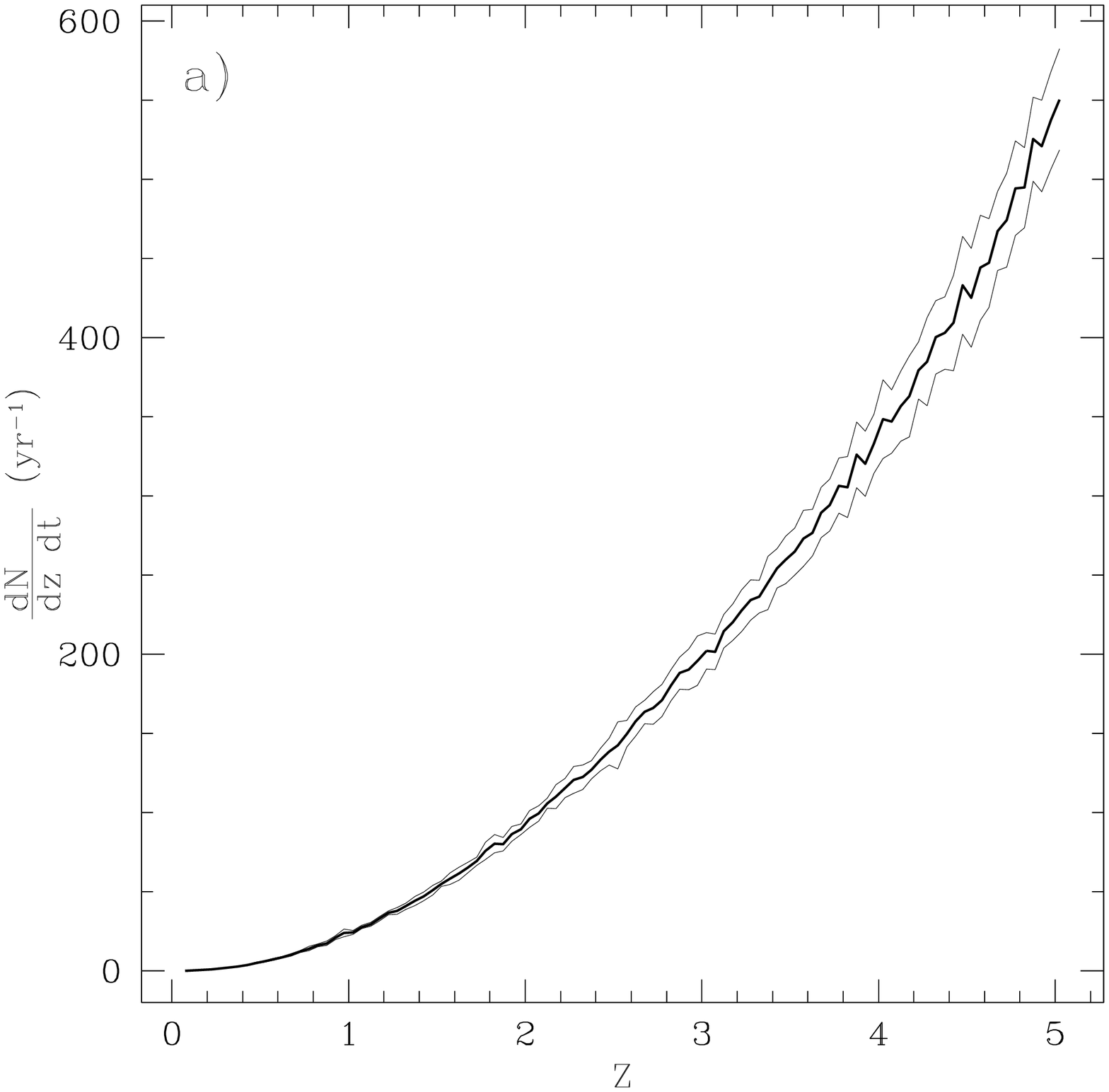}{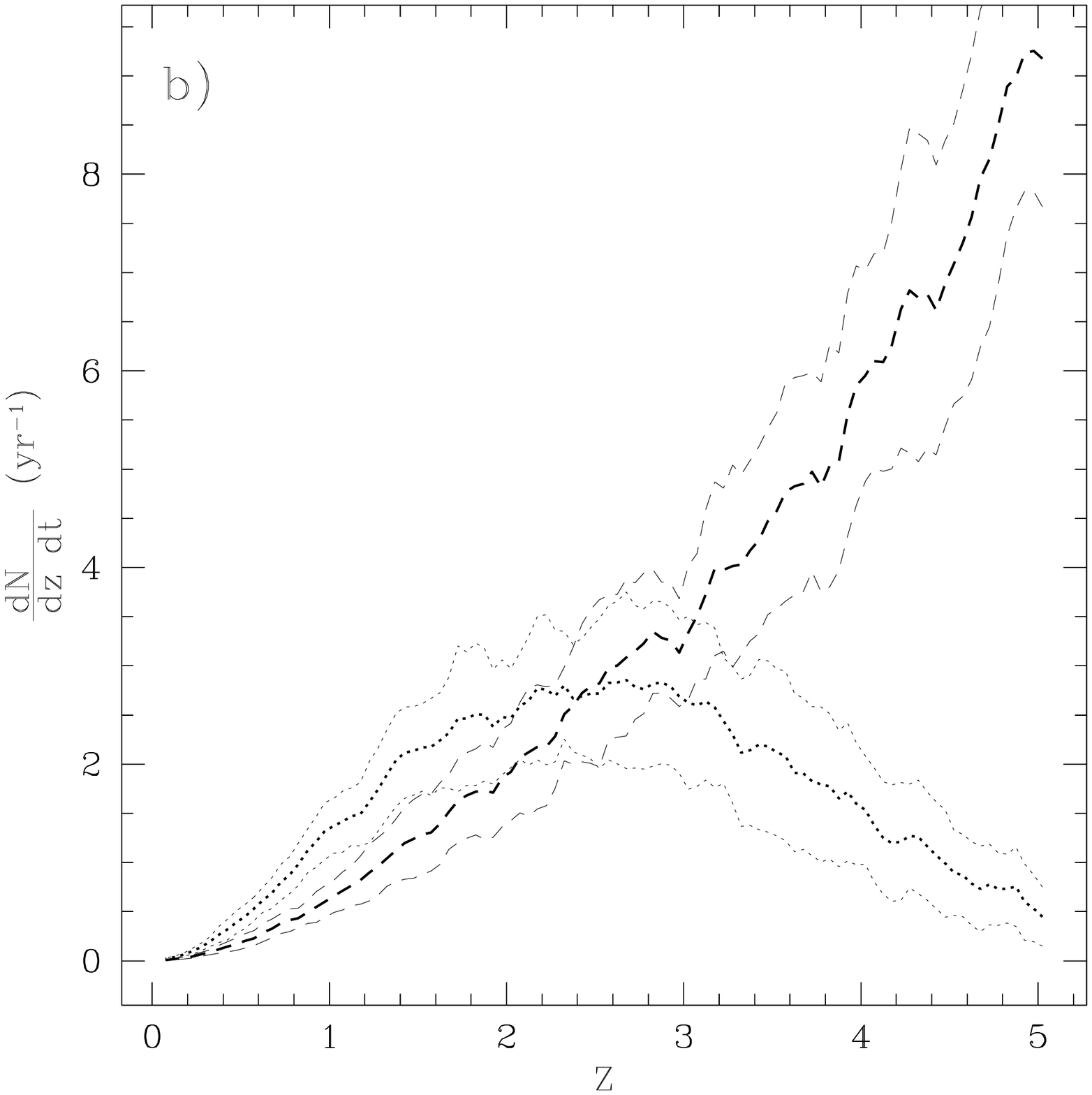}
\caption{Predictions for the black hole merger event rates (under the
assumption of very efficient black hole coalescence) in scenarios
where (a) all halos susceptible to harbor a black hole and (b) only a
fraction $f_{\rm BH}=10^{-2}$ of those, do contain a black hole. In
(b), the dotted line corresponds to model~I and the dashed line to
model~II; the event rates in (b) were smoothed over $\delta z=0.2$ for
better rendering. Standard deviations are indicated in each case by
the upper and lower thin lines. Various effects which could modify
these rates are discussed in the text.
\label{fig:two}}
\end{figure}

Figure~\ref{fig:two}b shows the event rates in model~I and model~II.
We also show in Figure~\ref{fig:two}a the event rates expected in a
third ``maximal'' model in which we assume that every single halo
described by the merger tree harbors a SMBH (i.e., $f_{\rm BHi}
=f_{\rm BH}=1$ at any redshift, independent of halo mass). The
predicted event rates differ both in terms of the total number of
events and in their distribution with redshift.

The maximal model shows that the merging rate of all the halos
described by the merger tree decreases steeply with cosmic time
(Fig.~\ref{fig:two}a). For the adopted cooling mass
(Eq.~[\ref{eq:coolmass}]), the merger tree predicts as much as $\sim
550$ merger events per year per unit redshift at $z=5$ and less than
$\sim 30$ merger events per year per unit redshift below $z=1$.  In
model~I, on the other hand, mergers of halos containing SMBHs are much
rarer because of the reduced population of SMBHs in halos ($f_{\rm
BHi}=3 \times 10^{-2}$, independent of mass). However, since mergers
gradually increase the value of $f_{\rm BH}$ with cosmic time
(Fig.~\ref{fig:one}a), the probability of a merger involving two SMBHs
initially increases with redshift.\footnote{We have checked that the
same model without the low--mass halos ``appearing'' from below our
mass resolution limit predicts essentially identical merging rates.}
The black hole merger event rate peaks at $z \sim 2.5$ where $\lsim 3$
events per year per unit redshift are predicted (Fig.~\ref{fig:two}b;
dotted line). This is the most pessimistic of the models considered
here, with a total event rate $\sim 9$~yr$^{-1}$ out to $z=5$.

In model~II, by construction, mergers of halos with SMBHs initially
occur only for halos more massive than $ \approx 5 \times 10^{9} {\rm
M_\odot}$. As halos merge together, this limit is increased further at
$z<5$, so that mergers occur on a typical mass scale which is growing
with cosmic time.  Halo merger rates generically peak at a
characteristic redshift that varies inversely with the halo mass scale
(i.e., the lower the halo mass, the higher the peak redshift); it
peaks at a redshift beyond $5$ for $\sim 10^{10} {\rm M_\odot}$ halos
(Lacey \& Cole 1993; see also Figure~5 in Haiman \& Menou 2000).  The
combination of these two properties leads to a decrease in the merger
rate with cosmic time in model~II (Fig.~\ref{fig:two}b; dashed line),
where the typical halo mass involved in a merger with SMBHs increases
with cosmic time. The merger tree predicts $\sim 9$ black hole merger
events per year per unit redshift at $z=5$, decreasing to less than
$\sim 0.5$ merger events per year per unit redshift below $z=1$. A
total of $\sim 15$ events per year out to $z=5$ is predicted.

\subsubsection{Low-Frequency Gravitational Waves}

SMBHs are prime targets for {\it LISA} (Folkner 1998) as sources of
low-frequency ($1-10^{-4}$~Hz) gravitational waves when they coalesce (see
Press \& Thorne 1972; Thorne 1987 and references therein). The merger tree
allows us to make more definite estimates of the black hole coalescence rates
than has been previously done (see Thorne \& Braginsky 1976; Haehnelt 1994;
1998).  These predictions must take into account the spectral coverage and
sensitivity of the instrument. However, following Haehnelt (1994; see also
Folkner 1998), we note that {\it LISA} should be able to detect the
gravitational signal from two coalescing SMBHs up to very high redshifts ($z >
10$) provided they both have masses $\gsim 10^3 {\rm M_\odot}$.

The masses of SMBHs are left unspecified in the merger tree
implementation described in \S2.  A simple ansatz for the SMBH masses
is to assume that they follow, up to $z=5$, the relation determined
locally:
\begin{equation}
M_{\rm BH} = 1.2 \times 10^8 {\rm M_\odot} \left( \frac{\sigma_e}{200~{\rm
km~s^{-1}}} \right) ^{3.75},
\label{eq:geb}
\end{equation}
where $\sigma_e$ is the velocity dispersion of the stellar bulge
(Gebhardt et al. 2000).  The relation between the stellar and dark
matter velocity dispersions can be specified based on the observed
radial stellar distributions ($\rho_{\rm stars}\propto r^{-3}$), and
assuming an isothermal DM halo ($\rho_{\rm DM} \propto r^{-2}$), which
yields $\sigma_e=(3/2)^{-1/2} \sigma_{\rm DM}$ (see, e.g., Turner,
Ostriker \& Gott 1984). Adopting this relation, we find that all the
SMBHs described by the merger tree have masses $> 10^3 {\rm M_\odot}$ (the
minimum SMBH mass described by the tree at $z=5$ is $\approx 1.3
\times 10^3 {\rm M_\odot}$, according to Eqns.~[\ref{eq:coolmass}]
and~[\ref{eq:geb}]).  In this case, all the merger events counted in
Figure~\ref{fig:two} are detectable by {\it LISA}.

{If instead of equation~(\ref{eq:geb}), we adopt the steeper relation 
of Merritt \& Ferrarese (2001a)
\begin{equation}
M_{\rm BH} = 1.3 \times 10^8 {\rm M_\odot} \left( \frac{\sigma_e}{200~{\rm
km~s^{-1}}} \right) ^{4.72},
\label{eq:fm}
\end{equation}
we find that our smallest halo harbors a SMBH with a mass $\lsim 100~M_\odot$
at $z= 5$, below {\it LISA}'s detection limit. In this case, the rate of events
detectable by {\it LISA} could be substantially reduced.  We note, however,
that both equations (\ref{eq:geb}) and (\ref{eq:fm}) have been obtained in
nearby galaxies.  In extending these relations to higher redshifts, one needs
to ensure that the total amount of mass in SMBHs is consistent with the
observed luminosity of the quasar population (Soltan 1982).  This implies that
in scenarios with rare SMBHs at high redshift, the typical SMBH in a halo with
a given mass will have to be more massive than in scenarios where SMBHs are
more widespread.  As a result, the above relations between velocity dispersion
and black hole mass derived locally should scale differently at high redshift,
making SMBHs more massive (and more detectable) in our models where $f_{\rm
BH}<1$. This suggests that only scenarios in which SMBHs are common at high
redshift will be affected by the above mass/sensitivity threshold for the rate
of merger events detectable by {\it LISA}.  This issue, which requires a
detailed accretion model for the SMBHs besides their merger history, will be
addressed in more detail elsewhere (Haiman, Menou \& Narayanan 2001, in
preparation).}

The gravitational wave signal provides a wealth of information on the
emission source (e.g. Thorne 1987). For coalescing black hole
binaries, the observables are the binary orbital period $P_{\rm orb}$
(twice the quadrupolar wave period), its time derivative $\dot P_{\rm
orb}$, and the gravitational wave amplitude $h$.  Modulo the unknown
binary inclination and source-detector orientation for individual
sources, one can put meaningful constraints on the binary mass and its
distance to the observer via a combination of these three observables
(e.g. Schutz 1986). Thus, the quantity $(h P_{\rm orb} / \dot P_{\rm
orb})$ provides a direct estimate of the distance to the source,
independent of its mass; the value of $h$ then constrains the mass of
the system once the distance is known. \footnote{In a cosmological
context, $P_{\rm orb}$ and $\dot P_{\rm orb}$ are affected by the
cosmological redshift and the distance becomes a luminosity distance,
but equally good constraints can be derived assuming the cosmology is
known (e.g. Schutz 1989).}  This raises the exciting possibility that
{\it LISA} could not just detect the events predicted in
Figure~\ref{fig:two}, but could also discriminate between coalescing
SMBHs at high and low redshift, constrain the masses involved in the
various merger events and, all together, probe the characteristics of
the population of SMBHs as a function of redshift.

\subsubsection{Black Hole Mergers}

The SMBH merger event rates shown in Figure~\ref{fig:two} were derived
assuming that SMBHs coalesce efficiently, i.e. shortly after the
merger of their parent halos (\S2.2).  Our current knowledge of the
merger process is limited, and various plausible physical effects
could substantially modify its efficiency.

In their seminal work, Begelman, Blandford \& Rees (1980) showed, for
a specific massive example, that the coalescence timescale for two
SMBHs in a galactic merger product could be very long, i.e.,
comparable to or larger than a Hubble time. Numerical simulations show
that dynamical friction between the SMBH and the surrounding star
cluster, while being an important effect before the phase of binary
evolution dominated by gravitational wave emission, is rather
inefficient, so that the coalescence of the two SMBHs is not an
obvious outcome (Rajagopal \& Romani 1995; Makino 1997; Quinlan \&
Hernquist 1997; Merritt, Cruz \& Milosavljevi\'c 2000). However, this
conclusion is complicated both by the possible acceleration of the
coalescence process due to the ``wandering'' of the SMBH binary in the
central stellar cluster --- a process that is difficult to follow
numerically (Quinlan \& Hernquist 1997; Merritt et al. 2000) --- and by
the possible influence of another merger on the preexisting stellar
distribution (Roos 1988; Polnarev \& Rees 1994; see also Begelman et
al. 1980 for a discussion of gas infall effects).

If SMBHs are brought together by successive halo mergers at a rate
higher than the rate at which they can coalesce, the lowest mass SMBH
is likely to be ejected out of the nucleus of the merger remnant by
the slingshot mechanism (Saslaw, Valtonen \& Aarseth 1974; Hut \& Rees
1992; Xu \& Ostriker 1994). This black hole ejection is more probable
in the maximal model presented above (where $f_{\rm BHi}=1$) than in
models~I and~II, where SMBHs are comparatively rare.  For example, in
the maximal model, a halo with mass $>10^{12} {\rm M_\odot}$ at $z=0$
typically has experienced about $10^2$ to $10^3$ mergers between
$0<z<5$ with other halos containing a SMBH. In fact, we find that in
the maximal model, $\sim 10\%$ of all the mergers occurring within a
redshift step $\Delta z=0.05$ (corresponding to $\approx 1.6 \times
10^7$~yrs at $z=5$ and $\approx 7 \times 10^8$~yrs at $z=0$ for the
adopted cosmology) are multiple mergers involving three or four halos
all of which contain a SMBH (indicating therefore very short times
between two successive mergers).  Note that most of the mergers happen
at high redshift in this model, as shown by Fig.~\ref{fig:two}a. In
models~I and~II, these ``multiple'' mergers within a redshift step
$\Delta z=0.05$ are exceedingly rare and the cumulative number of
mergers of halos with SMBHs involved in the building-up of the halos
at $z=0$ is much reduced. For example, halos of mass $>10^{11} {\rm
M_\odot}$ at $z=0$ have experienced up to $50$ mergers in model~I, and
up to $80$ mergers in model II, between $0<z<5$, with other halos
containing a SMBH (although the $\sim 20\%$ of halos that still do not
contain a SMBH at $z=0$ have not experienced any such merger; see
Fig.~\ref{fig:one}b).  In all three models, there is therefore the
possibility for slingshot ejections via three-body interactions,
unless the SMBH merging process is very efficient (i.e. coalescence
time $\ll$ than the typical time between successive mergers, at any
redshift).  Analytical arguments and numerical simulations, in the
context of dark matter halos being made of SMBHs, suggest that many of
the ejected SMBHs would remain undetectable and bound to their parent
galaxy, while some of them could escape to infinity (see, e.g., Hut \&
Rees 1992, Xu \& Ostriker 1994). This raises the possibility of the
existence of a population of free-floating SMBHs located outside
galaxies.

\section{Discussion}

Our description of the merger history of SMBHs is limited to redshifts
$z \leq 5$: we find that the merger tree fails to accurately reproduce
the PS mass function at higher redshifts (\S2.1).  In the model with a
SMBH in every halo satisfying our cooling criterion (the maximal
model), and in model~II, the halo merger event rates are strongly
peaked toward high $z$, suggesting even more events at $z >
5$. However, this simple extrapolation toward higher redshifts of
the expectations for {\it LISA}  is limited by the
tendency for SMBHs to be less massive at high $z$ (thereby moving the
coalescence signal out of the frequency range of {\it LISA}), and by
the fact that bound objects do not form beyond $z \sim 20-30$ in
popular CDM cosmologies (see Barkana \& Loeb 2001 for a recent
review).

Our models also suffer from the uncertainty on the minimum cooling
mass $M_{\rm min}$ (Eq.~[\ref{eq:coolmass}]). Because of the steep
slope of the halo mass function, most of the halos described by the
merger tree have masses close to $M_{\rm min}$, so that the
predictions for the merger event rates in the maximal model ($f_{\rm
BH}=1$) and, to a much lesser extent, in model~I are affected by the
value of $M_{\rm min}$ (more merger events are expected for smaller
$M_{\rm min}$).  Nevertheless, we expect our qualitative conclusion to
hold with different values of $M_{\rm min}$: (1) a few percent of the
massive halos at $z=5$ should harbor a SMBH to satisfy the constraints
at $z=0$ and (2) if SMBHs do not preferentially form between $3<z<5$,
then a fair fraction of luminous galaxies must contain a SMBH at $z
\sim 3$, the epoch of peak quasar activity.

We also note that the models~I and~II considered above are highly
idealized. In model~II, it is postulated that SMBHs form exclusively
in high-mass galaxies at $z>5$ and, implicitly, that no new SMBHs form
at $0<z<5$.  The assumption in model~I, namely that a small universal
fraction of the halos described contain SMBHs, could represent the
scenario envisioned by Eisenstein \& Loeb (1995), in which SMBHs form
only in halos with unusually low specific angular momentum. We note
that, in this scenario, halos in which SMBHs form should experience
fewer mergers because of their comparatively poor environments and
$f_{\rm BHi}$ should somewhat depend on the halo mass.  These two
effects are not included in our model~I.

As we have shown, the merging rates scale strongly with halo mass, so
that in a more realistic model the fraction $f_{\rm BH}$ would be an
increasing function of halo mass at almost any redshift (and in
particular at the arbitrary starting redshift $z=5$ chosen here)
because of previous mergers experienced by the halos at even larger
redshifts.  It is also possible that SMBHs continuously form with
cosmic time, with some complex dependence on redshift.  Nonetheless,
these idealized models represent two plausible extremes of the black
hole occupation functions, serving to illustrate the essentials of the
merging process and showing that SMBHs could be rare in halos at high
redshift without violating constraints at $z=0$.  Indeed, since the
same models with an additional, significant rate of SMBH formation
with cosmic time or an initial parameter $f_{\rm BHi}$ that is an
increasing function of halo mass would predict larger values of
$f_{\rm BHi}$ at $z=0$ for the large mass halos, the two models
explored here are useful extreme cases.

A source of uncertainty for the models presented here is that real
galaxies (e.g. those in clusters) can correspond to sub--halos of a
larger collapsed halo (e.g. the cluster).  This invalidates the
one-to-one correspondence between galaxies and Press--Schechter halos,
at least for galaxies residing in groups or clusters (see White,
Hernquist \& Springel 2001 for recent results on the multiplicity
function of sub--halos). We note that the sample of galaxies of
Magorrian et al. (1998), that is used here as the main observational
constraint for the models, is non--uniform, containing some galaxies
from the Virgo cluster, the Local Group, as well as other galaxies.
Implicit in our work is the assumption that nearly all galaxies in the
field (those with a one-to-one correspondence to Press--Schechter
halos) harbor SMBHs.  If future observations reveal that SMBHs are
less common in field galaxies, this could allow even more extreme
models than considered here (i.e. with even rarer SMBHs).

The results of Magorrian et al. (1998) have been quantitatively challenged, in
that the SMBH masses are now thought to be a factor of a few smaller than {
originally} estimated by Magorrian et al. (see, e.g., Merritt \& Ferrarese
2001b; Gebhardt et al. 2000).  {We note that these uncertainties do not
effect our conclusions here, since in our models, we only prescribe the
presence or absence of a SMBH. Although this has not been established to date,}
if SMBHs were found in the future to be rarer (not just less massive) than
estimated by Magorrian et al., the case for rare SMBHs at high redshift could
be made even stronger (i.e. the fraction of SMBHs populating galaxies at high
redshift could be even smaller than considered here).

{Evolution models with rare SMBHs can be further constrained by
requiring that they fit the observed luminosity function of quasars.
Since only a small fraction, $f_{\rm BH} \ll 1$, of high--redshift
halos are populated with SMBHs in these models, those halos which do
contain SMBHs had to harbor more massive SMBHs in these models than
they would have in standard evolution models with $f_{\rm BH}=1$.  In
standard models, the mass that constitutes a central SMBH at $z=0$ is
distributed among all the halo progenitors at some earlier redshift
$z$, each of whom harbors a SMBH.  In comparison, in models with rare
SMBHs, the same material resides at redshift $z$ in a fraction $\sim
f_{\rm BH}$ of the progenitors, so that the typical ratios of black
hole to halo mass are larger in models with rare SMBHs by a factor
$\sim f_{\rm BH}^{-1}$.  In standard models (see, e.g., Haiman \& Loeb
1998; Kauffmann \& Haehnelt 2000), the SMBH has a mass which is a
small fraction, $ \sim 10^{-2}-10^{-3}$, of the total gaseous mass of
the parent galaxy. This places an approximate limit of $f_{\rm
BH}\gsim 10^{-2}-10^{-3}$ on the maximum rarity of SMBHs.

Successful evolutionary models are also required to reproduce the
tight correlation between the SMBH mass and the stellar velocity
dispersion of the host galaxy bulge observed locally (Ferrarese \&
Merritt 2000; Gebhardt et al. 2000).  We emphasize that in the models
presented here, the masses of high-$z$ SMBHs or their accretion
history were not specified and that there is considerable freedom in
adjusting both of these quantities in an evolutionary model.  For
example, in the models of Kauffmann \& Haehnelt (2000) for the
evolution of active nuclei, the fraction of the gas that is accreted
onto the SMBH during a merger is chosen to scale with redshift.  We
expect that a suitable prescription for the accretion histories and/or
the initial SMBH masses can lead to a tight $M_{\rm BH}-\sigma$
relation as observed locally, and will pursue this question in a
future work.}

The above points demonstrate the need for a better characterization of
the population of SMBHs as a function of redshift. We highlighted the
possibilities and limitations of the detection of gravitational waves
associated with SMBH mergers for this purpose. It would be very useful
to test the extent of the population of SMBH via other means.  The
frequency of SMBH binaries at the centers of nearby galaxies may well
be different for the various models discussed above, although this
frequency depends on the efficiency with which SMBHs coalesce, and is
difficult to estimate theoretically.  A more promising test could be
to search for SMBHs at the centers of local dwarf galaxies with masses
smaller than currently probed, since our models predict that nuclear
SMBHs may not be ubiquitous in galaxies below a certain characteristic
mass scale at $z=0$. In the future, searches for nuclear activity at
high redshift will be useful in constraining models with rare SMBHs,
because they provide lower limits to the fraction of galaxies
harboring SMBHs (some of them being possibly inactive).

\section*{Acknowledgments}

We are grateful to Jeremy Goodman, David Spergel, Michael Strauss and
Scott Tremaine for useful discussions, and Martin Haehnelt for
comments on the manuscript.  Support for this work was provided by
NASA through Chandra Postdoctoral Fellowship grant number PF9-10006
awarded to KM by the Chandra X-ray Center, which is operated by the
Smithsonian Astrophysical Observatory for NASA under contract
NAS8-39073 and through the Hubble Fellowship grant HF-01119.01-99A,
awarded to ZH by the Space Telescope Science Institute, which is
operated by the Association of Universities for Research in Astronomy,
Inc., for NASA under contract NAS 5-26555.

\clearpage

\clearpage

\end{document}